# Visible-frequency metasurfaces for broadband anomalous reflection and high-efficiency spectrum splitting

*Zhongyang Li,[+] Edgar Palacios,[+] Serkan Butun and Koray Aydin[*]*

*Department of Electrical Engineering and Computer Science, Northwestern University,*

*Evanston, IL 60208*

[+]*Equally contributed to this work.*

[*]*Email: aydin@northwestern.edu*

**Abstract**

**Ultrathin metasurfaces have recently emerged as promising materials to enable novel, flat optical components and surface-confined, miniature photonic devices. However, experimental realization of high-performance metasurfaces at visible frequencies has been a significant challenge due to high plasmonic losses and difficulties in high-uniformity nanofabrication. Here, we propose a highly-efficient yet simple metasurface design comprising of single gradient antenna as unit cell. We demonstrate visible broadband (450 – 850 nm) anomalous reflection and spectrum splitting with 85% conversion efficiency. Average power ratio of anomalous reflection to the strongest diffraction was calculated to be ~ $10^3$ and measured to be ~ 10. The anomalous reflected photons and spectrum splitting performance have been visualized using CCD and characterized using angle-resolved measurement setup. Metasurface design proposed here is a clear departure from conventional metasurfaces utilizing multiple, anisotropic resonators, and could enable high-efficiency, broadband metasurfaces for achieving directional emitters, polarization/spectrum splitting surfaces for spectroscopy and photovoltaics.**

Controlling the flow and propagation of light has always been the main theme in the field of optics. Conventional optical devices such as prisms, lenses and waveplates operate based on the principles of geometrical optics and they are quite efficient in controlling and manipulating the light propagation through reflection, refraction or diffraction. However, such optical components are often bulky and offer limited performance in manipulating light-matter interactions at the wavelength or subwavelength scale due to diffraction limit. With recent advances achieved in the fields of nanophotonics and nanooptics, plasmonic nanostructures and metamaterials expand the range of matter's physical properties and facilitate substantial control of optical fields beyond the diffraction limit.[1,2]

More recently, metasurfaces,[3] metamaterials of reduced dimensionality, has been proposed and demonstrated to engineer phase retardation along a medium's surface, which opened up new opportunities towards achieving virtually flat optical components.[4,5] Such metasurfaces consist of space-variant optical resonators with a subwavelength size, whose optical response is designed to exhibit desired amplitude or phase shift through their geometry. Through an elaborate arrangement of arrays, metasurfaces have been previously shown to enable phase discontinuities and facilitate beam-steering, anomalous refraction and optical-wavefront shaping.[3] Metasurfaces have been designed to operate at visible,[6] infrared,[7,8] terahertz,[9] and microwave[10] frequencies. Various optical and photonic devices such as anomalous deflectors (reflectarray/transmitarray),[9,11-13] surface plasmon couplers,[14] vortex plates,[3,8] polarization converters,[15] flat lenses,[10,16-19] focusing mirrors,[20] waveplates,[21-23] and holograms,[6,24-26] and photonic spin-controllers[27,28] show the promise and potential of metasurfaces and their applications in various optical and photonic technologies.

Realization of visible frequency metasurfaces with broadband optical response and high efficiency is rather challenging due to subwavelength scale optical resonator elements (meta-atoms). A typical metasurface design utilizing several discrete resonator elements[3,4,7,9-11,14,20] offers limited conversion efficiency from incident beam to anomalous photons[7] due to the discrete multiple meta-atoms allowing regularly reflected/transmitted beams and other diffraction modes[3] in addition to ohmic losses.[17]

In this article, we propose and demonstrate a broadband anomalous meta-reflectarray across the visible frequency range and beyond, with high conversion efficiency by using single gradient "meta-atom" design as the building block. Comprised of trapezoid shaped nanorods, the flat metasurface is quite easy for fabrication with high geometric accuracy. The spatial-variant surface could impart a broadband gradient phase shift to outgoing photons without any cross-polarization effect and produce anomalous reflected rainbow components of red, orange, yellow, green and blue, which can be visually observed by naked eye or cameras. Such a quasi-continuous building block design with high duty cycle is capable of suppressing the normal reflection from the surface, greatly improving the conversion efficiency. High conversion efficiency, broadband visible metasurface suggests potential applications of high SNR optical spectrometer, polarization beam splitters,[7] high efficiency plasmon couplers,[14] directional emitter[29] and focusing lenses[10] and mirrors.[20]

A three-dimensional (3D) schematic of the meta-reflectarray is illustrated in Figure 1a. In contrast to the previous complex lattice designs containing multiple optical resonators, the proposed structure here is composed of an array with a single trapezoid-shaped Ag antenna as a building block on an optically-thick Ag film separated by a 50-nm-thick SiO$_2$ ($n = 1.45$) spacer. The trapezoid shaped Ag antenna arrays with the feature size of over 30 nm are precisely patterned on top of SiO$_2$ film using electron-beam lithography. The propagation direction of the illumination source is normally incident to the metasurface (along the *z*-axis) with orthogonal polarization along *x*-axis, as indicated in Figure 1a. Since the Ag film is optically thick (100 nm) to prevent light transmission *T*, the metasurface only allow reflection. The corresponding scanning electron microscope (SEM) image for the fabricated trapezoids is shown in Figure 1b.

Nanorod antenna are usually employed as prototype building block since it is the most straightforward and commonly studied plasmonic resonator.[1,30] Determined by geometry of the nanorod, it usually exhibits strong optical response from localized surface plasmon resonance (LSPR) when excited along the long side of nanorod, while short side has an off-resonance behavior. Here, we focus on the latter off-resonance mode for achieving high reflectance, hence low absorbance. 3D finite-difference time-domain (FDTD) simulations were performed using commercial software (Lumerical$^{TM}$) to calculate the optical response of nanorods. When excited along the short width, rectangular nanorods with uniform width of 30 - 150 nm reveal close-to-unity reflection amplitude and drastic interfacial phase shifts covering over 2π for the outgoing fields, as shown in Figure 1c. More discussion on reflection amplitude and absorptivity can be found in Figure S1 of Supplementary Information.

In order to achieve a broadband optical response, we intuitively propose a simple building block consisting of single trapezoid-shaped nanorod for designing meta-reflectarrays. By linearly changing its width from $W_s = 30$ nm to $W_l = 150$ nm, the relative phase shift along the metasurface is continuously modulated. Because the thickness of the trapezoid antenna is of deep-subwavelength scale ($\sim\lambda/20$), the phase accumulation over the light propagation path in such inhomogeneous medium can be neglected and the total phase discontinuity is only induced by the ultrathin metasurface.[6,7] Such introduction of interfacial abrupt phase discontinuity would redirect the "normal" reflection beam to an off-normal direction determined by the gradient of the phase change, which leads to the generalized Snell's law for reflection for normal illumination case.[3]

$$\theta_r = \arcsin\left(\frac{\lambda_o}{2\pi}\frac{d\Phi}{dy}\right) \quad (1)$$

The representative interfacial phase gradient curves (colorful dash line) are depicted in Figure 1c. Such gradient phase jump provides an additional effective wavevector to the reflected photons - along the *y*-axis, as schematically illustrated in Figure 1d. According to the generalized Snell's law, the theoretical calculation based on Equation (1) predicts that normal illumination at wavelength of 480 nm (blue), 520 nm (green), 560 nm (yellow) and 660 nm (red) will be steered with a reflection angle of 28.03°, 31.13°, 33.84° and 41.00°, respectively. It is worth noting that such anomalous reflection based on metasurfaces is conceptually different from the conventional blazed gratings with triangular, sawtooth-shaped grooves. Blazed gratings accumulate gradual linear phase delay through different length of light propagation path, essentially similar with oblique reflection from an inclined plane. The proposed metasurface arrays enable non-linear phase discontinuities at a flat 2D surface by excitation localized plasmonic modes from anisotropic resonators. Such propagation path reduced flat surface could provide new possibilities for integration into other photonic devices.

Most previous metasurface designs utilized multiple subwavelength-scale "meta-atoms" as their unit cells[3,4,7,9-11,14,20] producing discrete interfacial phase shifts, which limits the conversion efficiency and bandwidth operation. In stark contrast, the proposed single "meta-molecule" trapezoid-shaped antenna is continuously changing width (30 nm - 150 nm), which implies the interfacial phase shift $\Phi$ is a quasi-continuous function along the *y*-axis and leaves much less discrete space between unit cells. The higher duty cycle and quasi-continuous phase shift could strongly suppress the diffraction modes and significantly increase the power conversion efficiency[3] and operation bandwidth.

To experimentally characterize the performance of the meta-reflectarray, an angular resolved measurement setup was employed as schematically depicted in Figure 2a. A halogen lamp served as the broadband light source which was focused, collimated and polarized prior to reaching the metasurface sample at normal incidence. Reflected light from the sample was later collected at an angle $\theta_r$ with an Andor-Newton spectrometer system. Angle dependent reflection measurements were taken by maintaining the incident light normal to the sample and varying the collection angle. Anomalous rainbow reflection from the meta-reflectarray is visualized by taking a photograph, which can be directly observed by the naked eye or detected by camera and CCD, as shown in Figures 2b and 2c. Figure 2b is a photograph of the anomalous reflection from the meta-reflectarray under normal illumination during angular reflection measurements. The incident and reflected beam trace are captured by utilizing a moveable screen to scatter photons with long exposure. Figure 2c is also an actual photograph to visualize the propagation directions of the outgoing beams at a full range of reflection angle. The normal incident beam illuminates the metasurface sample through a centimeter-size aperture on a white paper screen and the reflected photons are redirected and projected back onto the paper screen. All directional reflections are categorized into three reflected beam modes: anomalous reflection mode (m = 1), normal reflection mode (m = 0) and first order diffraction mode (m = -1). The anomalous rainbow reflection light (m = 1) primarily propagates along positive reflection angles to the right-hand side. Minor reflected power from the first order grating diffraction effect is scattered to the left-band side (m = -1) while the weak normally reflected light transmits through the aperture (m = 0). In addition, since the meta-reflectarray design offers interfacial phase discontinuity only along the *y*-axis, polarization-dependent characteristics of the anomalous reflection could be observed, shown in the Supplementary Video and more discussion can be found in Supplementary Information.

Similar trapezoid patterns have been demonstrated to exhibit broadband high absorptivity in a recent study.[31,32] Here, we design the metasurface to exhibit low absorptivity and high reflectivity, in order to efficiently redirect incoming normal photons to the anomalous reflection direction. Our meta-reflectarray design exhibits high broadband reflectivity with an average of over 86% reflectivity between 450 – 800 nm (shown in Figure S2 of Supplementary Information). Figure 3a plots simulated angle-dependent reflection distribution at far-field from the metasurface that is normally illuminated with a broadband light source. The dash lines in Figure 3a indicate the three reflected beam modes which are normal reflection mode (m = 0), anomalous reflection mode (m = 1) and first order diffraction mode (m = -1) corresponding to Figure 2c. Most of the incoming power is reflected towards the anomalous direction. Reflected power for m = 0 and m = -1 modes are significantly lower than the anomalous reflection mode of m = 1. Specifically, the power ratio of anomalous reflection to normal reflection or diffraction mode is numerically calculated to be on the order of $10^3$ for visible bandwidth from 500 nm to 800 nm. The total conversion efficiency from incident illumination to anomalous reflection by the metasurface is calculated to be ~ 85% considering the intrinsic absorption. Comparing with conventional blazed gratings which usually show optimized efficiency only at specific wavelength, such planar metasurface based reflectarrays offer broader operation frequency with relatively higher conversion efficiency.

Figure 3b and 3c plot simulated directional reflected power spectra for m = -1 and m = 1 modes respectively, for detection angles between $\pm 30°$ to $\pm 52.5°$ with rotation interval of 2.5°.

Corresponding experimental data for m = -1 and m = 1 modes for same detection angles are plotted in Figures 3d and 3e. We observe that the measured spectra of directional anomalous reflections agree well with simulation results in terms of peak wavelength, bandwidth and relative power intensity. The bandwidth of reflected power spectra at a specific angle is as narrow as ~ 30 nm indicating low angular dispersion for reflected light. By comparing the intensities between m = 1 and m = -1 modes, it is clear that proposed meta-reflectarray can efficiently redirect photons to the anomalous reflection direction. The maximum ratio of anomalous reflection power (m = 1) to the diffraction order mode (m = -1) is calculated to be 1500 in simulation (Figure 3b-3c) and 15 in experiment (Figure 3d-3e).

In addition to the simulated and measured reflected power spectra, photographs of the metasurface arrays taken from different observation angles is provided in Figure 4a – 4d. Photographs has been taken by using a commercial camera from four different angles corresponding to 28 °, 31 °, 34 ° and 41 ° from the metasurface. One can easily observe the color change of blue, green, orange and red via naked eye or taking a photograph as a function of the reflection angle. We performed numerical simulations near the surface of the meta-reflectarray to understand how the optical power is redirected to the anomalous directions. Figure 4e-4h illustrates the calculated electric field profile of the reflected beam propagating away from the metasurface for four different wavelengths, 480 nm (blue), 520 nm (green), 560 nm (yellow) and 660 nm (red). Reflection angles are calculated to be 28 °, 31 °, 34 ° and 41 °, which agree very well with the measurement collection angles and calculated reflection angles based on Equation (1).

The anomalously reflected power distributions in polar coordinates are plotted in Figures 4i-4l for the corresponding wavelengths. The observed well-defined wavefronts of plane wave propagating along different reflection directions (Figure 4f-fh) could be confirmed by Figure 4i-4l that the other diffraction modes and normal reflection are strongly suppressed.

**Discussion**

In summary, we proposed and realized a broadband, visible frequency metasurface by employing single trapezoid plasmonic antenna as a building block. Our meta-reflectarray design offers easy fabrication, broadband operation and high conversion efficiency compared with conventional metasurfaces based on multiple, different size optical antennas. Through trapezoid "meta-molecule", we achieved quasi-continuous interfacial phase shift and significantly improved the conversion efficiency. Average power ratio of anomalous reflection to strongest diffraction mode was calculated to be on the order of $10^3$ and measured to be on the order of 10. Flat visible-frequency metasurfaces enable phase discontinuities along an optically flat surface based on localized plasmonic modes therefore provide new possibilities for integration into other photonic and plasmonic devices. In comparison with blazed diffraction gratings, our metasurface based reflectarray has the potential to operate with broader-bandwidth and higher conversion efficiency. The metasurface functionality suggests practical optoelectronic devices and applications for high SNR optical spectrometer, polarization beam splitters, light absorbers, high efficiency plasmon couplers, directional emitters as well as flat lenses and mirrors.

**METHODS**

**Metasurface Fabrication.** For the trapezoidal metasurface samples described here the substrate used was polished Si that was covered first with 100 nm of Ag followed by a 50 nm $SiO_2$ spacer using electron-beam evaporation. The patterned trapezoidal designs were then exposed in Bi-layer PMMA using a JEOL 9300 100kV electron-beam lithography system. The fabricated sample contains 10 identical nanostructured arrays with individual size of 200×200 μm$^2$, uniformly placed in a 5×2 matrix with total area of 1800×600 μm$^2$. To ensure highest resolution patterns development was completed using cold development at 5 °C with a solution of 7:3 IPA: H$_2$O ratio for 1min. The final layer, 30nm of Ag covered were deposited with electron-beam evaporation and lifted off in 1165 Microchem resist remover at 50 °C.

**Angular Measurement.** Experimental characterization of samples was completed by collection of reflection spectra using an angle resolved reflection system. A broadband halogen lamp was coupled into an optical fiber. Proper collimation of the light was found to be important in producing the sharpest reflection features for the desired wavelengths due to the design's high sensitivity to the incident angle. Thus, prior to reaching the sample, the light from the fiber was focused, collimated and polarized using a series of achromatic doublet lenses and a linear polarizer in the light pathway, which was helpful at normal incidence for all measurements. After focusing the 4X Nikon achromatic collection objective, the motorized rotation stage positon is manually adjusted to the eccentric point to eliminate image shift during data collection at multiple angles. The reflection is then coupled into a spectrometer consisting of a 303-mm-focal-length monochromator and Andor Newton electron multiplication charge-coupled device (EM-CCD) for a series of measurements with reflection angles between 30-52.5 ° at 2.5 ° intervals. Angular measurements at lower angles below 30 ° are restricted by the physical limitations of the setup as the collection lens start to block incoming photons. Measurements for angles that are above 52.5° is not reliable, since the optical signal from the metasurface is not enough to locate the position of the sample.

**Numerical Simulation.** Full-field electromagnetic wave calculations were performed using Lumerical$^{TM}$, a commercially available finite-difference time-domain (FDTD) simulation software package. 3D simulations for were performed for the proposed metasurface design with a unit cell area of 200×1000 nm$^2$ at *x-y* plane using periodic boundary conditions. Perfectly matched layers (PML) conditions are utilized along the propagation of electromagnetic waves (*z*-axis). Plane waves were normally incident to the nanostructures along the +*z* direction, and reflection and transmission is collected with power monitors placed behind the radiation source and after the structure, respectively. The reflected powers at a full range of angles are calculated by the far-field calculation option of the reflection power monitor. Electric and magnetic field distributions are detected by 2D field profile monitors in *x-z* plane. The complex refractive index of Ag for simulation is utilized from the data of Palik (0-2 μm) [33] and SiO2 is from the data of Palik[33].

**Acknowledgements**

This material is based upon work supported by the AFOSR under Award No. FA9550-12-1-0280. K.A. acknowledges financial support from the McCormick School of Engineering and Applied Sciences at Northwestern University and partial support from the Institute for Sustainability and Energy at Northwestern (ISEN) through ISEN Equipment and Booster Awards. This research was also partially supported by the Materials Research Science and Engineering Center (NSF-MRSEC) (DMR-1121262) of Northwestern University. This research made use of the NUANCE Center at Northwestern University, which is supported by NSF-NSEC, NSF-MRSEC, Keck Foundation, and the State of Illinois and the NUFAB cleanroom facility at Northwestern University. Use of the Center for Nanoscale Materials at Argonne National Laboratory was supported by the U. S. Department of Energy, Office of Science, Office of Basic Energy Sciences, under Contract No. DE-AC02-06CH11357. Z.L. gratefully acknowledges support from the Ryan Fellowship and the Northwestern University International Institute for Nanotechnology.

**Author Contributions**

Z.L and K.A. proposed the idea and designed the experiment, Z.L. performed numerical simulations and analytical modelling, E.P. and Z.L. fabricated the samples, E.P. and S.B. built the optical setup, E.P., Z.L and S.B. performed the measurements, Z.L., E.P. and K.A. wrote the manuscript. Z.L and E.P. contributed equally to this work.

**Competing financial interests**

The authors have no competing financial interests.
15

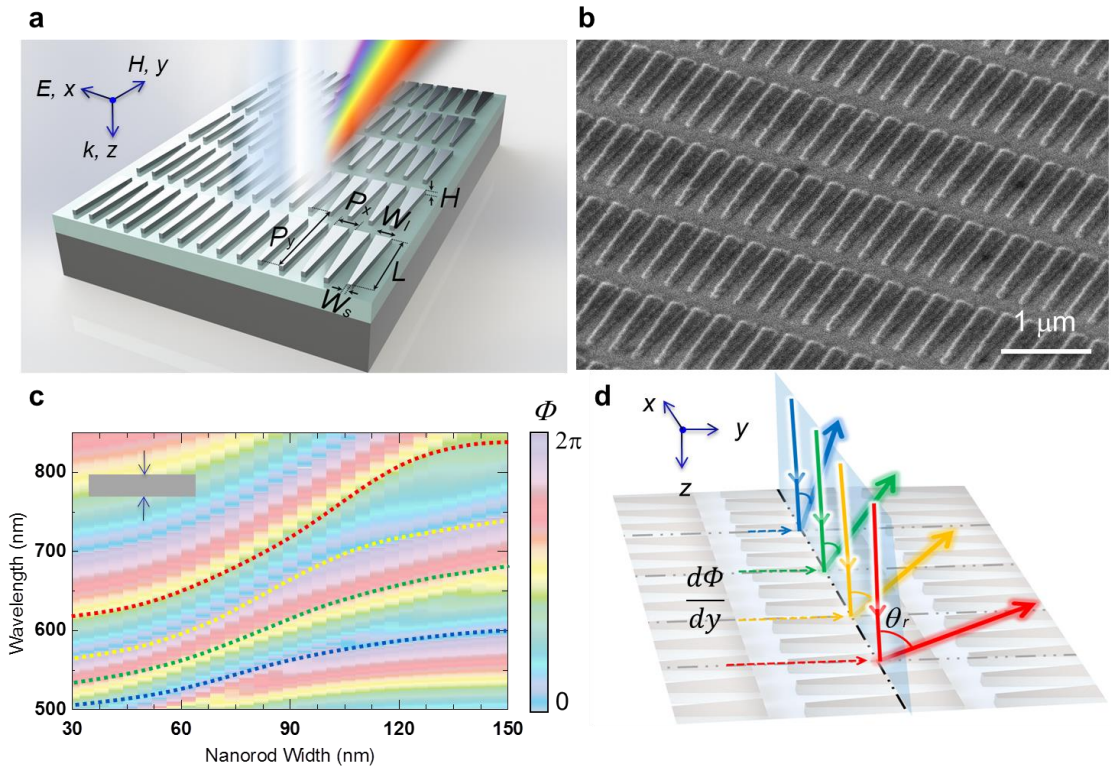

**Figure 1 Metasurfaces consisting of trapezoid shaped silver plasmonic antenna arrays and broadband gradient interfacial phase shifts for arbitrary bending reflection directions. a**, Schematic drawing of trapezoid-shaped nanorod array with geometric parameters of $P_x = 1000$ nm, $P_y = 200$ nm, $W_s = 30$ nm, $W_l = 150$ nm, $L = 800$ nm and $H = 30$ nm. Broadband white source is normal incident to the sample surface with the polarization along the *x*-axis. **b**, SEM image of the Ag trapezoid nanorods on top of $SiO_2$ film and Ag substrate. **c**, Simulated 2D map for phase shift of nanorods (inset) as a function of width and wavelength. The dash curves represent the gradients of phase shifts at different wavelengths. **d**, Schematics for illustrating that the gradient interfacial phase discontinuities $\frac{d\varphi}{dy}$ are varied for broadband source and provide different wavevectors along the surface to bend different wavelength components for reflecting into off-normal directions.

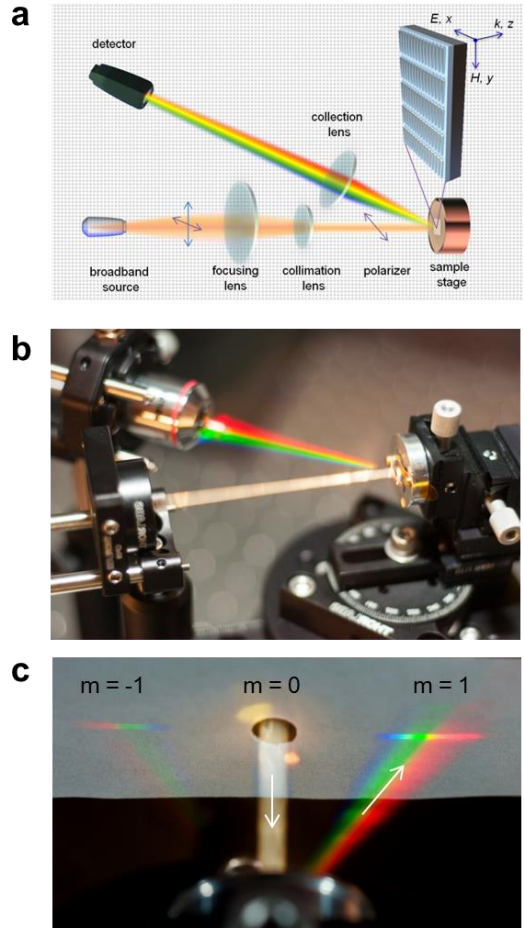

**Figure 2 Experimental characterization setup and actual photography for anomalous reflection. a,** Schematic drawing of angular reflection setup. **b,** Actual photograph for rainbow anomalous reflection beam redirected from normally incident broadband source by meta-reflectarray. The incident and reflected beam trace are realized by utilizing moveable screen to scatter photons with long exposure. **c**, Actual photograph for various reflection modes of redirected beam.

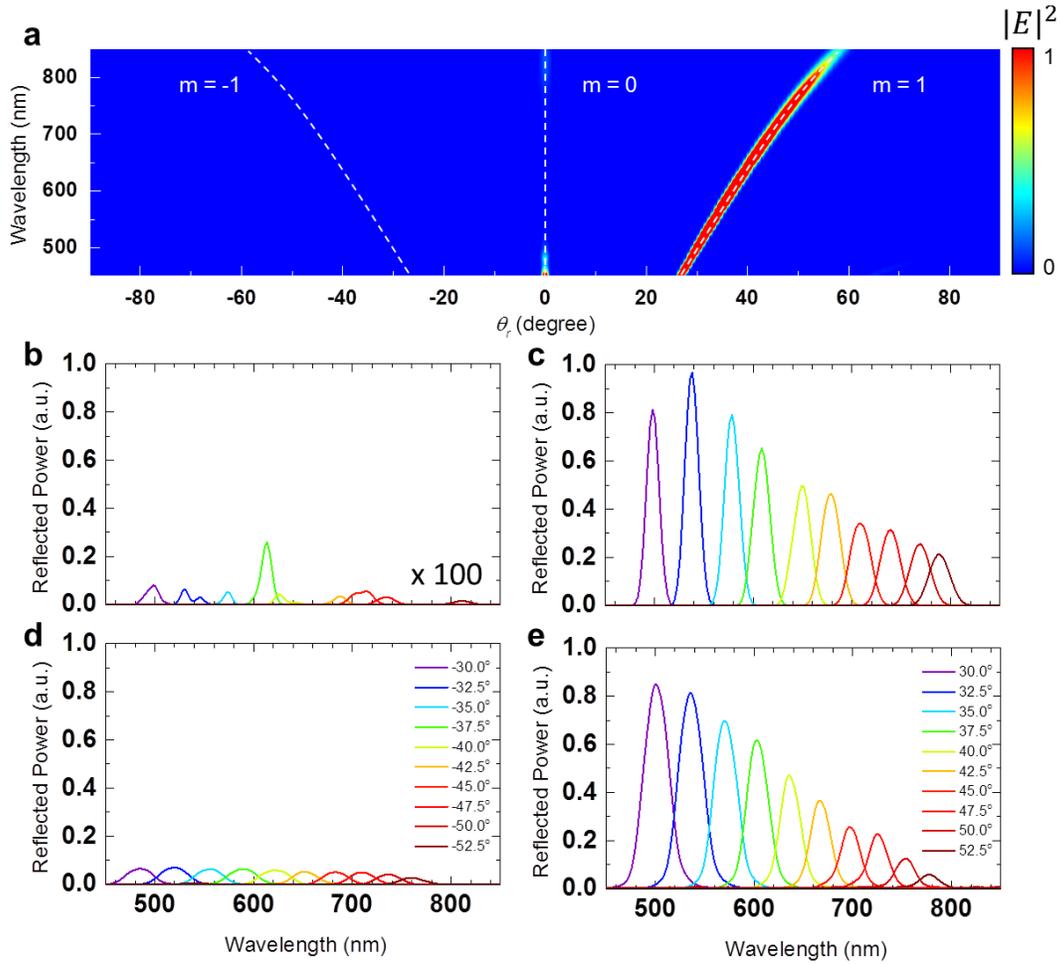

**Figure 3 Angular distribution of simulated and measured far-field reflection from the metasurface. a,** Simulated 2D contour for far-field reflected power as a function of reflection angles $\theta_r$ (x-axis) and wavelength (y-axis). **b,c,** Simulated reflected power spectra for different detection angles of -30°~ -52.5° and 30°~ 52.5°, respectively. The intensity of reflection power of **b** is multiplied by 100. **d,e,** Experimental reflected power spectra for different detection angles of -30°~ -52.5° and 30°~ 52.5°, respectively.

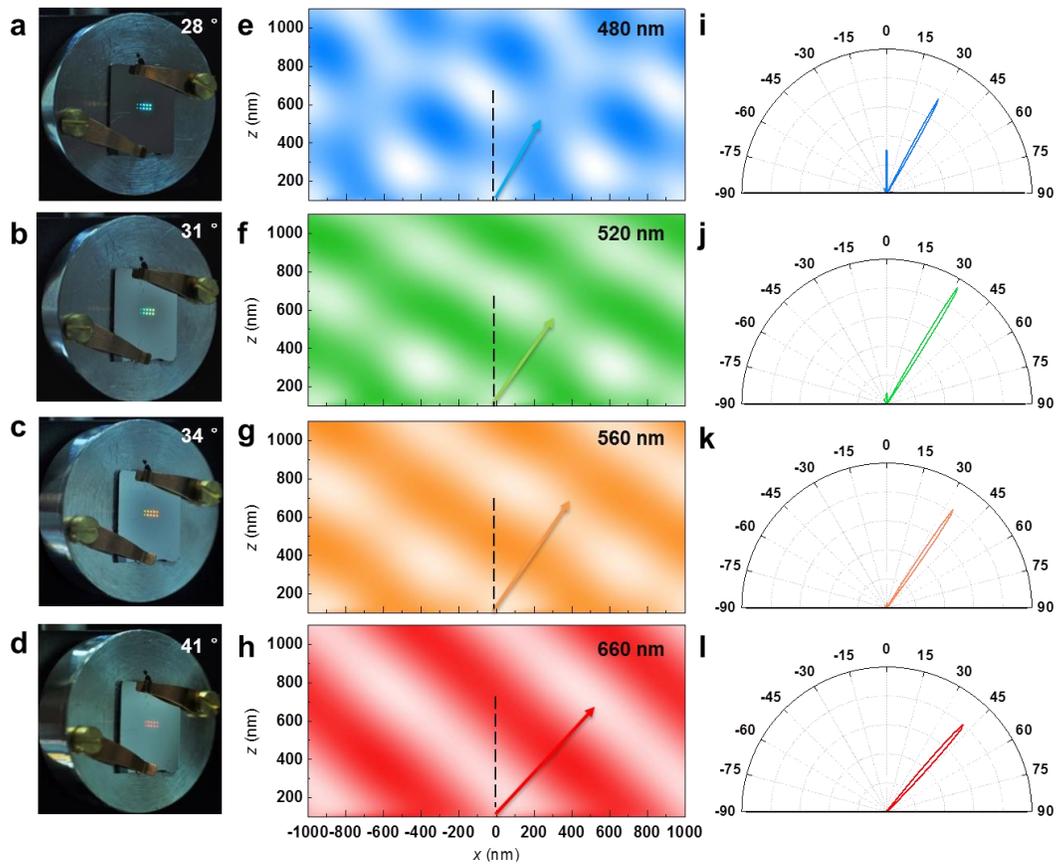

**Figure 4 Actual photography and near-field characterization on the metasurface by varying the detection angles. a-d**, Actual photographs for meta-reflectarray from different representative angles of 28 °, 31 °, 34 ° and 41 °. **e-h**, Calculated near-field profiles and wavefronts of reflected waves above metasurface at 480 nm, 520 nm, 560 nm and 660 nm, respectively. The arrows denote the propagation direction of the anomalous wave. **i-l**, Polar plots for far-field reflection spectra at corresponding wavelengths of **e-h**.